\documentclass[]{interact}

\usepackage[utf8]{inputenc}
\usepackage{epstopdf}
\usepackage[caption=false]{subfig}
\usepackage{amsmath, amssymb} 
\usepackage{booktabs}         
\usepackage{geometry}         
\usepackage{multirow}
\usepackage{algorithm}
\usepackage{algpseudocodex}

\usepackage{chngcntr} 
\usepackage{appendix}

\usepackage[numbers,sort&compress]{natbib} 

\bibpunct[, ]{[}{]}{,}{n}{,}{,} 
\renewcommand{\bibfont}{\fontsize{10}{12}\selectfont} 

\usepackage[colorlinks=true, linkcolor=blue, citecolor=blue, urlcolor=blue]{hyperref}
\hypersetup{pdfborder={0 0 0}} 

\theoremstyle{plain}

\theoremstyle{definition}

\theoremstyle{remark}

\begin{document}

\title{A Bayesian Non-linear Mixed-Effects Model for Accurate Detection of the Onset of Cognitive Decline in Longitudinal Aging Studies}

\author{
\name{F. Fernando Massa\textsuperscript{a}\thanks{CONTACT Fernando Massa. Email: fernando.massa@fcea.edu.uy} Marco Scavino\textsuperscript{a} and Graciela Muniz-Terrera\textsuperscript{b,c}}
\affil{\textsuperscript{a}Instituto de Estad\'istica, Departamento de M\'etodos Cuantitativos, Facultad de Ciencias Econ\'omicas y de Administraci\'on, Universidad de la Rep\'ublica, Montevideo, Uruguay; \textsuperscript{b}Heritage College Osteopathic Medicine, Ohio University, Athens, USA; \textsuperscript{c}Centre for Clinical Brain Sciences University of Edinburgh, Edinburgh, UK}
}

\maketitle

\begin{abstract}
Change-point models are frequently considered when modeling phenomena where a regime shift occurs at an unknown time. In ageing research, these models are commonly adopted to estimate of the onset of cognitive decline. Yet commonly used models present several limitations. Here, we present a Bayesian non-linear mixed-effects model based on a differential equation designed for longitudinal studies to overcome some limitations of classical change point models used in ageing research. We demonstrate the ability of the proposed model to avoid biases in estimates of the onset of cognitive impairment in a simulated study. Finally, the methodology presented in this work is illustrated by analysing results from memory tests from older adults who participated in the English Longitudinal Study of Ageing.
\end{abstract}


\section{Introduction}

In ageing research, determining the onset of cognitive decline is highly relevant since its accurate and early detection allows a better understanding of the ageing process, its characteristics, and factors associated with this onset \cite{karr2018}. Early detection is therefore critical for the implementation of preventive or therapeutic interventions that can slow or mitigate cognitive decline. The inaccurate estimation of this onset can have significant and potentially harmful consequences, for instance, overestimation of the onset of cognitive decline may result in older adults not receiving timely care and support. On the other hand, underestimating the onset of cognitive decline can lead to unnecessary anxiety and worry. Individuals may think they are in a more advanced state of cognitive deterioration than they are, negatively impacting their quality of life and emotional well-being. Therefore, there is a need for methods that permit the accurate estimation of the onset of cognitive decline in older adults, backed by solid data and appropriate clinical assessments.\\
Change-point (CP) models are commonly used in aging research for estimating the onset of cognitive decline \cite{hall2000, muniz-terrera2011, hout2011} and assessing other research questions \cite{sprague2020}. The detection of the moment at which a CP occurs in the trajectory of a stochastic process is a problem that has been addressed from multiple perspectives. It is a common problem in time series analysis \cite{aminikhanghahiS2017} and of utmost interest in longitudinal studies, where a set of individuals is followed over time. Although both types of studies aim to obtain predictions about the behaviour of paths of stochastic processes, longitudinal studies usually pay more attention to the determinants of the phenomenon under investigation. In this sense, statistical analysis is frequently performed in the context of regression models, often including random effects. Change-point regression models are commonly formulated in longitudinal studies within a framework based on linear mixed models \cite{kiuchi1995, mclain2014, muggeo2008} and more specifically, in the field of the study of the ageing process \cite{domenicus2008, hout2013, yu2012}. Commonly used model specifications include a change over time that may well be abrupt, such as in the \emph{Broken stick model} (BSM) \cite{cohen2008}, or gradual, as in the \emph{Bacon \& Watts model} (BWM) \cite{bacon1971} and the \emph{Bent cable regression model} (BCR) \cite{chiu2006}.\\
Formulating models suitable for longitudinal data within a DE framework relaxes linearity assumptions like those imposed in linear mixed-effects models. Regarding the explicative features of cognitive decline, the use of mixed models within a non-linear setting maintains the advantages of this methodology, such as describing population and individual variation and accounting for dependent data, while also allowing the incorporation of aspects concerning the onset of the decline phase and the factors associated with its delay or advancement, as well as model the speed at which this process occurs. We introduce a Bayesian non-linear mixed effects model in which the longitudinal trajectory is modeled through a differential equation (DE). DE models are increasingly used in longitudinal studies \cite{albano2012, hu2019, rosenstroem2013}, allowing the representation of complex dynamics where the passage of time plays a fundamental role. This novel research contribution focuses on aspects such as describing the temporal evolution of a specific phenomenon and predicting future observations. It is also possible to consider the mean of a longitudinal mixed-effects model as a particular case of a linear DE. \\
The rest of the paper is organized as follows. In Section~\ref{sec2} we review three CP regression models that are often used in practice, and then focus on the new DE model we propose. Statistical inference for these models is presented under a Bayesian framework, and we provide criteria for the model selection stage. In Section~\ref{sec3} we describe a simulation study designed to evaluate the performance of the DE model under different data-generating processes (DGPs) comparing it against the models presented in Section~\ref{sec2}. An application of cognitive data from the English Longitudinal Study of Ageing (ELSA) \cite{steptoe2013} is illustrated in Section~\ref{sec4}, where the proposed DE model shows superior prediction accuracy than the other three models and, on average, shifts the estimate of the onset of the cognitive decline two years later. Finally, in Section~\ref{sec5} we draw conclusions and propose future lines of research.

\section{Materials and methods}\label{sec2}
We formulate CP regression models within a non-linear mixed modeling framework (NLMM) as described by \cite{demidenko2013}. Assuming that the $n_i$ measurements of the $i$-th individual are contained on the vector $Y_i=(y_{i1},y_{i2},\ldots,y_{in_i})$, we adopt the hierarchical definition of NLMM expressed through the following equations:
\begin{subequations} \label{subeqnexample}
	\begin{equation}
		y_{ij}=f(t_{ij}, \theta_{0i}, \ldots, , \theta_{Ki}) + \varepsilon_{ij}, \qquad i=1,\ldots,n \quad j=1,\ldots,n_i,
		\label{eq:NLMMa}
	\end{equation}
		\begin{equation}
		\theta_{ki} = \beta_{k0}+\eta_{ki}, \qquad k=1,\ldots,K,
		\label{eq:NLMMb}
	\end{equation}
\end{subequations}
where $\varepsilon_{ij}\sim N(0,\sigma_{\epsilon}^2)$ and $\eta_{ki}\sim N(0,\omega_{k}^2)$.\\
Under this specification, the outcome $y_{ij}$ (response of subject $i$ at time $t_{ij}$) is a non-linear function of time and a set of parameters $\theta_{0i}, \ldots,  \theta_{Ki}$. Equation \ref{eq:NLMMa} is often referred to as the ``individual-level model''. Additionally, the parameters of this equation may include random effects $\eta_{ki}$. Equation \ref{eq:NLMMb} is often referred to as the ``population-level model''. In addition, both random effects ($\eta_{ki}$) and model errors ($\epsilon_{ij}$) are assumed to follow a normal distribution with zero mean and variance parameters $\omega_{k}^2$ and $\sigma_{\epsilon}^2$ respectively. An advantage of the NLMM lies in its ability to make predictions about the future values of a particular individual or an average individual.\\
The most commonly used CP models presented in the literature, as well as our new proposal, are introduced below. All the models are formulated within an NLMM framework by only considering an intercept $\beta_{k0}$ for each parameter in the populational equation. However, the formulation of models with explanatory variables can be easily generalized.\\

\subsection{Change point models}
Several formulations of CP models have been used in the literature. Here we recall the BSM as a piecewise linear model with one free knot. It is characterized by the following piecewise-continuous linear function:
\begin{equation}
	f_{BSM}(t,\boldsymbol\theta) = \left\{\begin{array}{ll}
		\theta_0+\theta_1 t, &t\leq \theta_{CP} \\
		\theta_0+\theta_1\theta_{CP}+\theta_2(t-\theta_{CP}) t, &t> \theta_{CP} \,.\\
		\end{array}
		\right.
	\label{eq:bsm}
\end{equation}
Another characterization has been provided in the BWM with a smooth transition function given as:
\begin{equation}
	f_{BWM}(t,\boldsymbol\theta) = \theta_0+\theta_1(t-\theta_{CP})+\theta_2(t-\theta_{CP})\tanh\left(\frac{t-\theta_{CP}}{\theta_T}\right)
	\label{eq:bwm}
\end{equation}
where tanh(.) denotes the hyperbolic tangent function.\\
Finally, we consider the BCR where two linear functions are connected by a quadratic polynomial:
\begin{equation}
	f_{BCR}(t,\boldsymbol\theta) = \left\{\begin{array}{ll}
		\theta_0+\theta_1 t, &t\leq \theta_{CP}-\theta_{T} \\
		\theta_0+\theta_1\theta_{CP}+\theta_2\frac{(t-\theta_{CP}+\theta_T)^2}{4\theta_T}, &\theta_{CP}-\theta_{T} < t \leq \theta_{CP}+\theta_{T}\\
		\theta_0+(\theta_1+\theta_2)t-\theta_2\theta_{CP}, &t> \theta_{CP}+\theta_{T} \,. \\
		\end{array}
		\right.
	\label{eq:bcr}
\end{equation}
Including random effects and explanatory variables in these models is optional; this choice reflects which parameter is allowed to vary among individuals, as presented in equation \ref{eq:NLMMa}. Although the interpretation of the intercept $(\theta_0)$ and slope parameters $(\theta_1,\theta_2)$ is not the same across models, the parameter associated with the CP $(\theta_CP)$ has the same meaning in all three models. Equations \ref{eq:bwm} and \ref{eq:bcr} propose smoother alternatives to the abrupt change of the BSM, where the smoothness of such change is controlled by the value of the transition parameter $(\theta_T)$ and a smooth function (hyperbolic tangent and a quadratic polynomial respectively).

\subsection{Differential equation model (DEM)}\label{sec2_2}
\begin{figure}[h]
	\centering
	\includegraphics[width=0.5\textwidth]{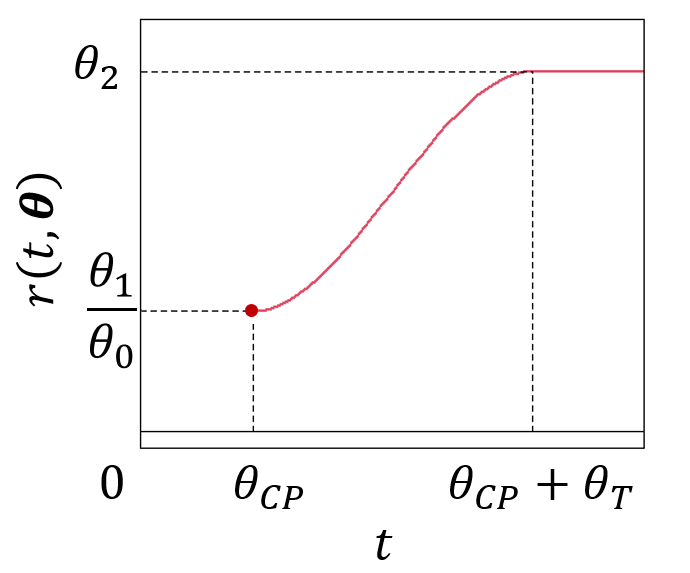}
	\caption{Rate function for the DEM.}\label{rate_function}
\end{figure}

Unlike the three previous alternatives, the model presented in equation \ref{eq:dem} is based on the description of a rate of change that is not constant over the course of ageing. The mean trajectory is described by a simple exponential decay DE where the key element of this model is the rate function $r(t,\boldsymbol\theta)$:
\begin{equation}
	f_{DEM}'(t,\boldsymbol\theta) = \left\{\begin{array}{ll}
	  \theta_1, &\quad t \leq \theta_{CP} \\
		r(t,\boldsymbol\theta)f_{DEM}(t,\boldsymbol\theta), &\quad t \geq \theta_{CP} \\
		f_{DEM}(\theta_{CP},\boldsymbol\theta)=\theta_0 \,.&\\
		\end{array}
		\right.
	\label{eq:dem}
\end{equation}
The family of solutions obtained by solving this simple DE is determined from the specification of the rate function $r(t,\theta)$. In this study, a non-decreasing rate function as presented in Figure~\ref{rate_function} is proposed. This specification of the rate function resembles the mean function corresponding to the BCR model where two straight lines are connected by a polynomial. However, the obtained mean function poses a different alternative.
Up to the point where the deterioration process begins, the rate of change is zero, which translates into a horizontal trajectory of the $f$ function at the value $\theta_0$. Then there is a transition period where the rate begins to increase up to a maximum value where it stabilizes. From this moment on, the decay is proportional to the cognitive state. Equation \ref{eq:transition} presents the components of the rate function:
\begin{equation}
	r(t,\boldsymbol\theta) = \left\{\begin{array}{ll}
		\theta_1/\theta_0, &t= \theta_{CP} \\
		p_3(t), &\theta_{CP} < t \leq \theta_{CP}+\theta_{T}\\
		\theta_2, &t> \theta_{CP}+\theta_{T} \,.\\
		\end{array}
		\right.
	\label{eq:transition}
\end{equation}
As in the previous cases, the CP is modeled through the parameter $\theta_{CP}$. Additionally, a transition period (of length $\theta_T$) around the CP is considered. Lastly, $p_3(t)$ is a polynomial of third grade that smoothly connects both parts of the rate function. To this end, it enforces the following constraints:
\begin{itemize}
  \item $p_3(\theta_{CP})=\theta_1/\theta_0$
	\item $p_3'(\theta_{CP})=0$
  \item $p_3(\theta_{CP}+\theta_T)=\theta_2$
	\item $p_3'(\theta_{CP}+\theta_T)=0 \,.$
\end{itemize}
The coefficients of this polynomial can be found solving a linear system (see Appendix~\ref{app_A}). Under this specification, it is possible to obtain the closed analytical expression for the $f_{DEM}(.)$ function presented next in equation~\ref{eq:el_dem}:
\begin{equation}
		f_{DEM}(t,\boldsymbol\theta) = \left\{\begin{array}{ll}
	  \theta_0+\theta_1 \left(t-\theta_{CP}\right), &\qquad t \leq \theta_{CP} \\
		\theta_0e^{-\int_{\theta_{CP}}^{t}r(s,\boldsymbol\theta)ds}, &\qquad t > \theta_{CP} \,. \\
		\end{array}
		\right.
	\label{eq:el_dem}
\end{equation}
The expression $\theta_0-\theta_1\theta_{CP}$ could be thought as an intercept like parameter for the linear segment before the CP. Applying straightforward algebra it is possible to expand the exponential expression in \ref{eq:el_dem} as follows:\\
\begin{equation}
		\left\{\begin{array}{ll}
	  \theta_0e^{-\mathcal{P}_3(t)}, &\qquad \theta_{CP} < t \leq \theta_{CP}+\theta_{T} \\
		\theta_0e^{-\mathcal{P}_3(\theta_{CP}+\theta_{T})+\theta_2(t-\theta_{CP}-\theta_{T})}, &\qquad t > \theta_{CP}+\theta_{T} \,.\\
		\end{array}
		\right.
	\label{eq:el_exp}
\end{equation}
Being $\mathcal{P}_3(t)=\int_{\theta_{CP}}^{t}p_3(s)$. Additionally, by redefining $\bar{\theta}_0=\theta_0e^{-\mathcal{P}_3(\theta_{CP}+\theta_{T})}$, the translated exponential decay appears more clearly.\\
Nonetheless, we prefer to present the model by means of equations \ref{eq:dem} and \ref{eq:transition} since they provide a clearer interpretation of the parameters.\\

\begin{figure}[h]
	\centering
	\includegraphics[width=0.75\textwidth]{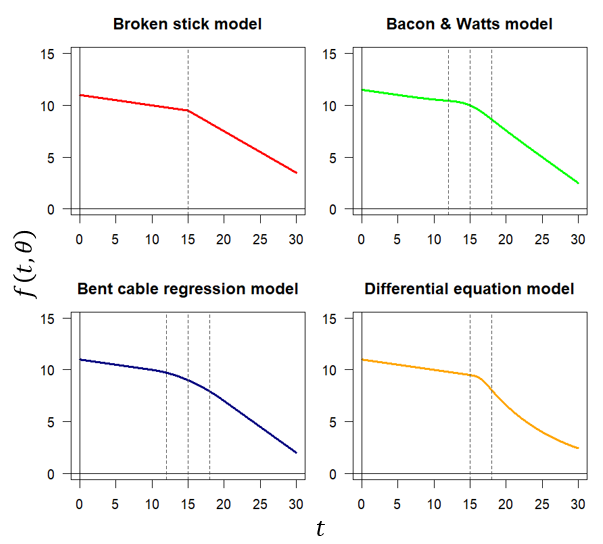}
	\caption{Change point models.}\label{models}
\end{figure}

To better understand how these models work, Figure~\ref{models} compares the mean trajectory of the four alternatives presented in equations \ref{eq:bsm}-\ref{eq:dem}.
Figure~\ref{models} shows how the four models consider three phases in the mean function. The BSM consists of two lineal segments with different slopes abruptly joined at the CP. The other alternatives adhere to this pattern, adding a ``transition'' phase between.\\
Regarding the transition parameter, it has different meanings in the four models. In the BWM its meaning is not trivial, but a ``radius of curvature'' (see page 528 of \cite{bacon1971}) can be constructed around the CP allowing for a smoother transition between both regimes. The BSM can be viewed as a limiting case of the BWM when the transition parameter equals zero, resulting in an abrupt transition. In the case of the BCR, the transition parameter represents the semi-amplitude of the transition period around the CP. Finally, in the DEM, it is the amplitude of the period that connects both the lineal and the decay phases. This period starts at the CP and ends at beginning of the decline phase.\\
The DEM has the advantage that the value of the function $f_{DEM}$ is never less than zero and, as seen in Figure~\ref{models}, is the only specification capable of capturing a non-linear behavior after the CP. The latter is of relevant since, as will be presented in scenarios based on Monte Carlo simulation, an inappropriate choice at CP model selection could result in significant biases in estimating the CP.

\subsection{Bayesian inference}
Due to the inclusion of random effects, a CP and the specification of a transition parameter, the proposed DEM belongs to the family of non-linear models and requires iterative and computationally demanding estimation methods. Hence, we opted for the use of a Bayesian estimation approach, that can better handle a (possibly) large number of random effects without resorting to numerical methods to solve high-dimensional integrals \cite{lee2022}. As we will see below, the estimation techniques were based on efficient Markov chain Monte Carlo (MCMC) algorithms.\\
We selected the following prior distributions. We assigned non-informative gaussian priors to the fixed effects with precision parameters set to 0.001. On the other hand, for the CP and the transition parameter we propose uniform priors whose range accounted for the duration of the study. Finally, as suggested by \cite{gelman2006, polson2012} we assigned half-Cauchy distributions as a weakly informative prior for the parameters associated with the error and random effects variances. These choices allowed us to write the models described above as in Figure~\ref{bayes}:
\begin{figure}[h]
	\centering
    \includegraphics[width=0.75\textwidth]{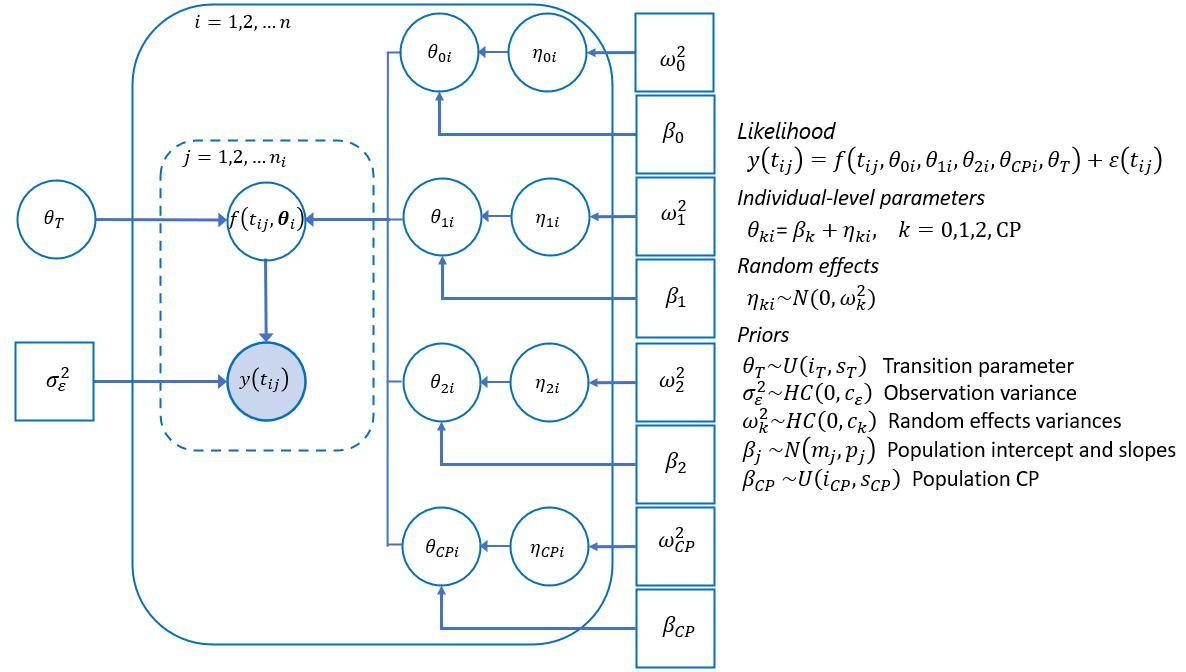}
	\caption{Graphical model of the hierarchical Bayesian non-linear mixed model. Nodes represent variables of interest (observed=shaded, latent=unshaded), with dependencies represented via the graph structure.}\label{bayes}
\end{figure}
In all cases, we run four parallel Markov chains of model parameters for 5000 iterations each. After discarding the first 2500 iterations of each chain, we assess the convergence of the MCMC algorithm with the four separate Markov chain samples of size 2500. Convergence was monitored using the $\hat{R}$ statistic, trace-plots and the effective sample size \cite{gelman2013}. All statistical analyses were conducted using R \cite{rcoreteam2024} making use of the MCMC algorithms available in the rstan and rjags libraries \cite{rstan,rjags}.

\subsection{Model selection}
One way to perform model selection within a Bayesian framework is by using the marginal likelihood \cite{llorente2012}. In model selection, $K$ competing models are considered and researchers are interested in the relative plausibility of each model $\mathcal{M}_k\,, k \in 1,2,\ldots,K\,,$ given the priors and the data. This relative plausibility is contained in the posterior model probability $p(\mathcal{M}_k | y)$ of model $\mathcal{M}_i$ given the data:
\begin{equation}
p(\mathcal{M}_k | y) = \frac{p(y | \mathcal{M}_k) \, p(\mathcal{M}_k)}{\sum_{k=1}^{K} p(y | \mathcal{M}_k) \, p(\mathcal{M}_k)}
    \label{eq:py}
\end{equation}
where $p(y | \mathcal{M}_k)$ is the marginal likelihood of the data under the $\mathcal{M}_k$ model. Thus, from the initial representation of modeling uncertainty contained in $p(\theta_k | \mathcal{M}_k)$ and $p(\mathcal{M}_k)$, the posterior distribution of the model $p(\mathcal{M}_k | y)$ updates this uncertainty quantification after observing the data. For the applications considered in this study, the calculations necessary to compute these quantities were obtained using the algorithms of the bridgesampling R library \cite{gronau2020}.\\
In addition, another extensively used indicator is the widely applicable information criteria (WAIC) \cite{watanabe2010}: 
\begin{equation}
    \text{WAIC} = -2\left\{\displaystyle\sum_{i=1}^n\log\mathbb{E}_{post}\left[p(y_i|\theta)\right]-\displaystyle\sum_{i=1}^n\mathbb{V}ar_{post}\left[\log p(y_i|\theta)\right]\right\}
\end{equation}
where $\mathbb{E}_{post}$ and $\mathbb{V}ar_{post}$ represent the mean and variance with respect to the posterior distribution $p(\theta|y)$.
Unlike posterior model probabilities, WAIC aims to assess the accuracy of out-of-sample predictions. To this end, it uses the log-predictive density (which is a more general quantity than the mean squared error) and a bias correction that take into account the effective number of parameters. For the purposes of this study, the algorithms used to obtain the WAIC were those contained in the loo R library \cite{vehtari2017}.\\
The fit indicators described above were introduced to assess the performance of the new DEM against BSM, BWM and BCR. In this paper, the emphasis is on the better estimation of the CP. For this reason, the comparison is conducted in terms of bias and interval coverage. Lastly, to explore the overall fit of each model, the posterior model probabilities as well as the WAIC value are provided.\\
To accomplish this, we proposed three simulation-based experiments. The aim of these experiments is to compare the fit of the models in different situations and to determine their performance to estimate the CP. Finally, the performance of the DEM versus the other competitors is presented using real cognitive data from ELSA.

\section{Simulation study}\label{sec3}
We designed a simulation study based on three experiments. The first has the DEM as the data-generating process (DGP). In this case, not only is the DEM expected to present the best fit, but it is also expected to find negative biases in the other three model specifications when estimating the CP. The second scenario corresponds to a situation where the DGP is the BSM. We considered this scenario to assess how robust the DEM is when it is not the appropriate model. Finally, the third scenario is similar to the first one, but limiting the follow-up period after the CP so that the curvature of the trajectory is not sufficiently decisive to point to the DEM as the best model.\\
In all scenarios, we simulated 1000 data sets composed of 50 individuals that were observed on 10 occasions at random at times between 0 and 20. The value of the parameter $\beta_1$ is set to 0 since the model is designed for application in studies of cognitive decline, where it is natural to assume that the cognitive state of the individuals remains constant until the moment when cognitive decline begins. For this reason, each scenario considered a horizontal trajectory up to the CP, which was fixed in the middle of the follow-up period $(\beta_{CP}=10)$. In Scenario 2, the slope after the CP $(\beta_2)$ was set at -0.5, while scenarios 1 and 3, considered a decline rate of -0.5. In each case, the value of $\beta_0$ was set at 11, the observation noise had a variance $\sigma^2_{\varepsilon}$ of 1.4, the random effects had variances of 0.3, 0.1 and 2 for the intercept, the slope (or rate) and the CP, respectively. Finally, the transition parameter ($\theta_T$) was set to 3. The algorithm to generate the data sets follows the next pseudo-code.

\begin{algorithm}
    \caption{Data simulation for the s-th scenario}
    \label{algo:alg1}
    \begin{algorithmic}[1]
        \Require{$N=200; n_{obs}=50; n_{ij}=7; T_{max}=20;$ DGP parameters}
        \Ensure{$N$ data sets}
        \Statex
				\For{$k \gets 1$ to $N$}
					\For{$i \gets 1$ to $n_{obs}$}
					\begin{enumerate}
						\item Using a uniform distribution, generate 10 random times between 0, and $T_{max}$ ($t_{i1},t_{i2},\ldots,t_{i10}$).
						\item Generate individual parameters from population-level equation~\ref{eq:NLMMb} using inputs and simulated random effects (see Figure 3) $\theta_{0i},\theta_{2i},\theta_{CPi}$.
						\item Generate observations $y_{i1},y_{i2},\ldots,y_{i7}$ using the individual-level equation~\ref{eq:NLMMa}.
					\end{enumerate}
				\EndFor
        \EndFor
    \end{algorithmic}
\end{algorithm}

In each scenario, the four models considered were fitted to each of the 1000 data sets using the hierarchical Bayesian non-linear mixed model proposed in Figure~\ref{bayes}. The choices of prior distributions could be observed in Appendix~\ref{app_B} (see Table B1). Information on the posterior distribution of the parameters was extracted after verifying the convergence of the MCMC algorithm (see the distribution of the $\hat{R}$ statistic and $n_{eff}$ for the $\hat{\beta}_{CP}$ parameter in Figure~\ref{fig:mcmc_diag} in Appendix~\ref{app_C}). The posterior median (PM) of the CP was used as the point estimator and credibility intervals (CrI) were constructed using the 0.025 and 0.975 sample quantiles (higher posterior density intervals were also constructed but did not differ significantly from those reported in Table~\ref{table_1}). The performance of the four models in estimating CP was evaluated through the bias of the $\beta_{CP}$ estimate, calculated as the difference between the real value and its corresponding estimate. It was considered that a successful (unbiased) estimation should include zero within the CrI of the bias. Likewise, using the CrI of $\beta_{CP}$ obtained in each of the 1000 data sets (as well as the true value of the CP), the effective coverage of this parameter was approximated in all models for each scenario. In this case, it is desirable that these values be as close as possible to 95\%. Additionally, the values obtained from the posterior distribution of the parameters of each model were used to calculate posterior model probabilities and WAIC (and its standard error). We followed the next pseudo-code to obtain estimates, intervals, and fit indicators from every data set. 

\begin{algorithm}
    \caption{Data simulation for the sth scenario}
    \label{algo:alg2}
    \begin{algorithmic}[1]
        \Require{$N$ simulated data sets}
        \Ensure{bias, effective coverage, and posterior probabilities}
        \Statex
				\For{$k \gets 1$ to $N$}
				\begin{enumerate}
					\item Fit $BSM$, $BWM$, $BCR$ and $DEM$ to data set $k$.
					\item Get PM and 95\% CrI of CP.
					\item Compute fit indicators.
					\end{enumerate}
        \EndFor
        \Statex Summarize values obtained in c).
    \end{algorithmic}
\end{algorithm}

Table~\ref{table_1} presents the PM and the 95\% CrI for the CP parameter, its bias, and the effective coverage probability of CrI. Finally, posterior model probability and WAIC are presented for the four models in each scenario.
\begin{table}[h]
    \centering
    \begin{tabular}{lccc}
    \hline
            & Scenario 1 & Scenario 2 & Scenario 3 \\
    \hline
    DGP & DEM & BSM & DEM \\
    \hline
    \multicolumn{4}{l}{\textit{CP Estimate (95\% CrI)}} \\
    \hline
    \quad BSM & 8.89 (7.96 ; 9.84) & 9.98 (8.42 ; 11.56) & 10.06 (9.31 ; 10.91) \\
    \quad BWM & 9.23 (7.84 ; 16.13) & 11.86 (8.65 ; 13.38) & 10.49 (9.08 ; 18.78) \\
    \quad BCR & 9.20 (7.97 ; 15.58) & 10.91 (9.11 ; 13.82) & 9.99 (9.34 ; 10.91) \\
    \quad DEM & 10.40 (9.68 ; 11.24) & 10.06 (8.50 ; 11.67) & 10.56 (9.45 ; 14.17) \\
    \hline
    \multicolumn{4}{l}{\textit{CP Bias (95\% CrI)}} \\
    \hline
    \quad BSM & -1.11 (-2.04 ; -0.16) & -0.02 (-1.58 ; 1.56) & 0.06 (-0.69 ; 0.91) \\
    \quad BWM & -0.77 (-2.16 ; 6.13) & 1.86 (-1.35 ; 3.38) & 0.49 (-0.92 ; 8.78) \\
    \quad BCR & -0.80 (-2.03 ; 5.58) & 0.91 (-0.89 ; 3.82) & -0.01 (-0.66 ; 0.91) \\
    \quad DEM & 0.40 (-0.32 ; 1.24) & 0.06 (-1.50 ; 1.67) & 0.56 (-0.55 ; 4.17) \\
    \hline
    \multicolumn{4}{l}{\textit{CP Effective Coverage (95\% CrI)}} \\
    \hline
    \quad BSM & 0.25 (0.18 ; 0.31) & 0.92 (0.89 ; 0.96) & 0.946 (0.92 ; 0.98) \\
    \quad BWM & 0.18 (0.13 ; 0.23) & 0.79 (0.73 ; 0.84) & 0.817 (0.76 ; 0.87) \\
    \quad BCR & 0.25 (0.19 ; 0.31) & 0.90 (0.86 ; 0.94) & 0.952 (0.92 ; 0.98) \\
    \quad DEM & 0.92 (0.88 ; 0.96) & 0.95 (0.92 ; 0.98) & 0.849 (0.80 ; 0.90) \\
    \hline
    \multicolumn{4}{l}{\textit{Posterior Model Probability}} \\
    \hline
    \quad BSM & \textless{}0.001 & 0.25 & 0.09 \\
    \quad BWM & \textless{}0.001 & 0.19 & 0.11 \\
    \quad BCR & \textless{}0.001 & 0.26 & 0.07 \\
    \quad DEM & \textgreater{}0.99 & 0.31 & 0.73 \\
    \hline
    \multicolumn{4}{l}{\textit{WAIC (SE)}} \\
    \hline
    \quad BSM & 1841.83 (30.50) & 1651.82 (29.38) & 1670.53 (30.85) \\
    \quad BWM & 1829.44 (30.31) & 1653.00 (29.37) & 1670.10 (30.70) \\
    \quad BCR & 1842.73 (30.29) & 1652.69 (29.44) & 1671.69 (30.85) \\
    \quad DEM & 1666.15 (26.61) & 1658.81 (29.31) & 1657.94 (30.34) \\
    \hline
    \end{tabular}
    \caption{Results from the simulation experiments.}
    \label{table_1}
\end{table}

The results presented in Table~\ref{table_1} suggest that the DEM has a superior performance when the decline after the CP shows enough curvature (Scenario 1). This can be seen in the posterior model probability, the lowest value of the WAIC estimate, on the unbiased nature of the CP estimator (the credible interval for the bias covers the value zero without being too wide), and on the effective posterior coverage probability of the credible interval of the true CP being very close to the nominal value of 95\%. Furthermore, the performance of the BSM is suboptimal, as it exhibits a negative bias in the estimation of the CP.\\
In addition, it can be observed that in the other two scenarios, the DEM performance competes with the other alternatives, even in the worst case (Scenario 2). Both if the true decline is linear (Scenario 2) or close to linear (Scenario 3), the CP estimated value not only is unbiased but also has nominal posterior coverage close to the true value. It is worth noting that even when the post-CP curvature is not sufficiently pronounced (Scenario 3), the posterior model probability indicates moderate evidence in favor of DEM.

\section{Application}\label{sec4}
\subsection{Data}
The English Longitudinal Study of Ageing (ELSA) is a prospective, population-based cohort of individuals aged over 50 to understand different aspects of ageing in England \cite{steptoe2013}. The cohort was established in 2002 with data collected biannually. To date, there have been nine waves of monitoring, in which interviews have been conducted with 19221 individuals. We focus on a memory marker, which is called the total word recall test \cite{mcfall2013}. This was constructed as the sum of both immediate and delayed recall tests. Participants were presented with a list of ten words and prompted to remember them, recall them immediately, and after approximately five minutes. This test serves as a measure of verbal skills and working memory.

\subsection{Results}
\begin{figure}[h]
    \centering
    \includegraphics{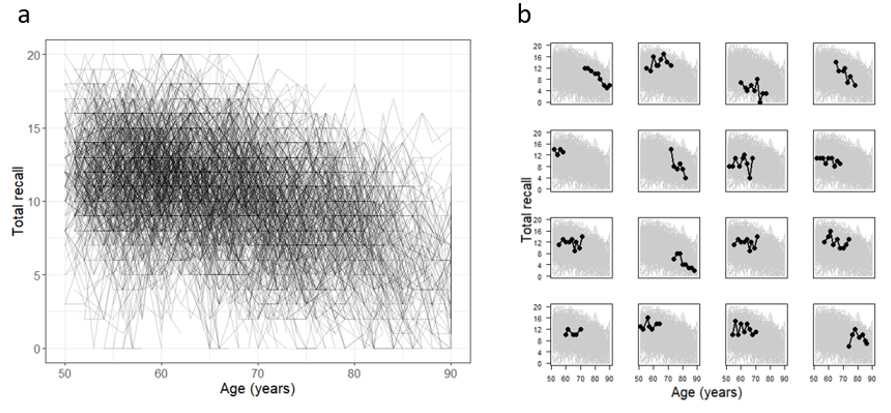}
    \caption{a.5000 randomly selected trajectories of total recall. b. A set of 16 individual trajectories showing that individuals not only may have a different number of measurements but also participate in the study at different periods.}\label{total_recall}
\end{figure}

Figure~\ref{total_recall}a presents a spaghetti plot of the total recall test score versus age over a sample of 5000 randomly selected participants. Additionally, Figure~\ref{total_recall}b displays some highlighted trajectories in black over the light gray trajectories from the first panel. Although Figure~\ref{total_recall}a suggests significant variability among trajectories, a ``stable'' phase followed by a ``downward'' phase beginning between 60 and 70 years of age can be discerned. This pattern is also evident in Figure~\ref{total_recall}b, where individual trajectories vary in level but consistently show that the rate of change differs between the beginning and the end of the follow-up period.

We fitted the four models introduced in the change-point models Section and calculated the model selection criteria presented in the model assessment Section. Table~\ref{table_2} presents the estimated parameters, including posterior medians and credible intervals, as well as model selection indicators for the three baseline models: BSM, BWM, BCR, and the proposed DEM. 

\begin{table}[H]
    \centering
    \begin{tabular}{lcccc}
    \hline
                                    & BSM             & BWM             & BCR             & DEM            \\ \hline
    \multicolumn{5}{l}{\textit{Fixed Effects}}                                                            \\ \hline
    \multirow{2}{*}{$\beta_0$}      & 11.29           & 11.28           & 11.31           & 11.22          \\
                                    & (11.19 ; 11.40) & (10.17 ; 11.39) & (11.20 ; 11.42) & (11.11 ; 11.32) \\
    \multirow{2}{*}{$\beta_2$}      & -0.26           & -0.12           & -0.26           & -0.05          \\
                                    & (-0.29 ; -0.24) & (-0.14 ; -0.11) & (-0.28 ; -0.23) & (-0.06 ; -0.04) \\
    \multirow{2}{*}{$\beta_{CP}$}   & 68.61           & 67.77           & 68.33           & 74.28          \\
                                    & (67.30 ; 70.50) & (66.14 ; 69.20) & (66.85 ; 69.94) & (72.93 ; 75.60) \\
    \multirow{2}{*}{$\theta_T$}     &                 & 3.33            & 2.47            & 3.63           \\
                                    &                 & (0.26 ; 4.95)   & (0.13 ; 4.82)   & (0.69 ; 4.96)   \\ 
    \hline
    \multicolumn{5}{l}{\textit{Random Effects}}                                                           \\ \hline
    \multirow{2}{*}{$\sigma_{\epsilon}$}   & 2.27          & 2.27          & 2.27          & 2.24           \\
                                           & (2.24 ; 2.30) & (2.25 ; 2.30) & (2.25 ; 2.30) & (2.22 ; 2.27)  \\
    \multirow{2}{*}{$\sigma_{\beta_0}$}    & 2.12          & 2.13          & 2.13          & 2.16           \\
                                           & (2.05 ; 2.20) & (2.05 ; 2.20) & (2.06 ; 2.20) & (2.08 ; 2.32)  \\
    \multirow{2}{*}{$\sigma_{\beta_2}$}    & 0.11          & 0.06          & 0.11          & 0.03              \\
                                           & (0.09 ; 0.13) & (0;05; 0.07) &  (0.09 ; 0.14) & (0.02 ; 0.04)  \\
	\multirow{2}{*}{$\sigma_{\beta_{CP}}$} & 6.84          & 6.68          & 6.74          & 8.39           \\
                                           & (6.05 ; 7.80) & (5.81 ; 7.49) & (5.86 ; 7.61) & (7.74 ; 9.09)  \\ 
    \hline
    \multicolumn{5}{l}{\textit{Model Selection}}                                                          \\ \hline
    WAIC                                  & 1233242.5     & 123265.3      & 123223.6      & 122827.6       \\
    PMP                                   & 0.229         & 0.228         & $\leq$0.001   & 0.542          \\
    \hline
    \end{tabular}
    \caption{Parameter estimates, 95\% credible intervals, and model selection indicators.}
    \label{table_2}
\end{table}

\begin{figure}[H]
    \centering
    \includegraphics[width=0.8\textwidth]{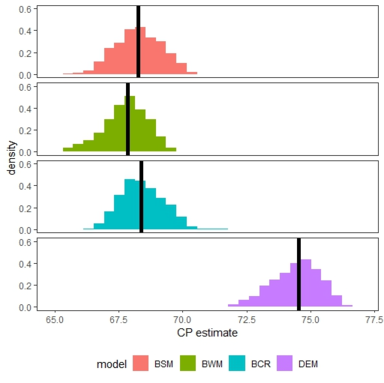}
    \caption{Posterior CP distribution according to the four proposed models.}
    \label{changepoints}
\end{figure}

On the one hand, all four models agree on the average intercept and the random components estimates. However, although the posterior model probability provides shows only anecdotal evidence in favor of DEM, the value of WAIC favors it as well. As shown in Figure~\ref{changepoints}, the DEM indicates a CP around 6 years later than those fitted by the other models. 

As seen in the simulation experiments, this could be due to non-linearities on the mean trajectory. Additionally, the DEM presents a narrower transition period, which is compensated by a higher variability in the random component of the CP.

\begin{figure}[H]
    \centering
    \includegraphics{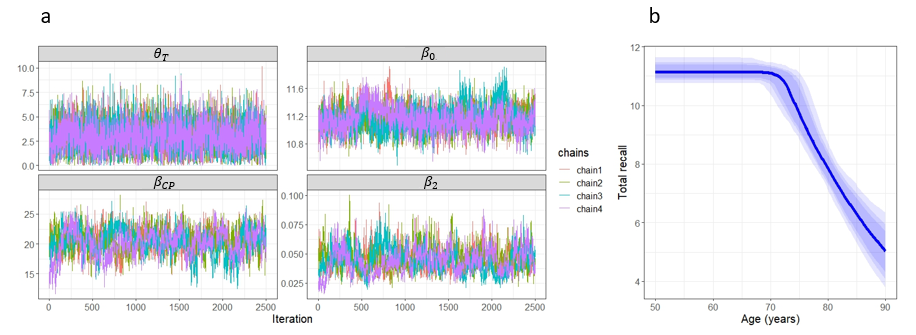}
    \caption{a. Trace plots from $\beta_0$, $\beta_2$,$\beta_{CP}$ and $\beta_T$. b. Fitted trajectory of total recall according to the DEM.}\label{fitted_total_recall}
\end{figure}

Finally, using the simulated values from the posterior distribution of the parameters from the DEM, Figure~\ref{fitted_total_recall} presents trace plots from the fixed effects of the DEM and the fitted curve along a 95\% probability band for total recall.

\section{Conclusions}\label{sec5}
This paper introduced a non-linear alternative based on a new DEM specification to commonly used CP models and presented its performance compared to the models most commonly used in the literature. Model fitting was carried out in a Bayesian context within the R framework using the rjags or rstan libraries. Rjags has proven to be faster than rstan fitting the models considered in this work. Nevertheless, parameter estimation could also be undertaken straightforwardly within a frequentist framework. Comparison among models was conducted by computing model selection indicators that rely on the MCMC samples.\\
Different simulation scenarios were implemented to explore the performance of the DEM with reference to the BSM, BWM and BCR models. From the analysis of the results of the simulation design, we observed that when the cognitive decline phase had a sufficiently pronounced curvature, the DEM presented the best fit and produced a less biased estimator for the CP with higher posterior coverage in the credible interval. However, when there were not enough observations in the period where the curvature manifests itself, the performance of the DEM decreased. Furthermore, when the DGP did not exhibit curvature in the post-CP phase, the performance of DEM was found to be on par with that of its competitors.\\
Ultimately, in the illustration of the models, we used actual cognitive data from the ELSA study. The results showed a slightly better fit of the DEM, which suggested an onset of the cognitive decline stage an average of 2 years later than the other models. This result is of particular interest in the area of ageing because it provides vital information for health policy planning.
In conclusion, it is worth noting that the model described in this paper is not meant to replace the models proposed in the literature, but rather to serve as a viable alternative, offering commendable statistical properties, transparent parameter interpretation, and sensible biological features (such as the mean function consistently avoiding intersection with the x-axis). About its use, it should be preferred when there are reasons to include non-linear behavior in the modeling stage. 
Future research could extend the new DEM by including explanatory variables, consideration of missing data, software development, and new applications of the DEM in other disciplines.

\section{Disclosure statement}
The authors declared that they have no potential conflicts of interest in relation to the research, authorship, and/or publication of this article.
\section{Data and code availability}
To obtain ELSA data from all waves, including wave 0 (Health Survey from England) contact the UK Data Service. The code used to generate the simulated data sets and obtain the results presented in this paper can be obtained by request to the main author.

\begin{appendix}

\section{Coefficients of $p_3(t)$}\label{app_A}
The polynomial in the rate function $r(t,\boldsymbol{\theta})$ has the form:
\[a_0+a_1t+a_2t^2+a_3t^3.\]
Imposing the constraints presented in Section~\ref{sec2_2}, the following system of linear equations is obtained:\\

\begin{equation}
\begin{bmatrix} 
1 & \theta_{CP} & \theta_{CP}^2 & \theta_{CP}^3\\
0 & 1 & \theta_{CP} & \theta_{CP}^2 \\
1 & \theta_{CP}+\theta_{T} & \left(\theta_{CP}+\theta_{T}\right)^2 & \left(\theta_{CP}+\theta_{T}\right)^3\\
0 & 1 & 2\left(\theta_{CP}+\theta_T\right) & 3\left(\theta_{CP}+\theta_{T}\right)^2 
\end{bmatrix}\left[ \begin{array}{c} a_0 \\ a_1 \\a_2 \\ a_3 \end{array} \right] = \left[ \begin{array}{c} \theta_1/\theta_0 \\ 0 \\ \theta_2 \\ 0\end{array} \right] \,.
\end{equation}

The coefficients of $p_3(t)$ that arise after solving this system are:
\begin{itemize}
  \item[] $a_3=-2\left(\theta_1/\theta_0-\theta_2\right)/\theta_T^3$
  \item[] $a_2=3\left(2\theta_{CP}+\theta_{T}\right)\left(\theta_1/\theta_0-\theta_2\right)/\theta_T^3$
  \item[] $a_1=-6\theta_{CP}\left(\theta_{CP}+\theta_{T}\right)\left(\theta_1/\theta_0-\theta_2\right)/\theta_T^3$
  \item[] $a_0=\theta_1/\theta_0-\left(\theta_1/\theta_0-\theta_2\right)\theta^2_{CP}\left(2\theta_{CP}+3\theta_{T}\right)\theta_T^3 \,.$
\end{itemize}

\section{Prior distributions}\label{app_B}

\setcounter{table}{0}
\renewcommand{\thetable}{B\arabic{table}}

Table~\ref{priors_table} provides a detailed description of the selected a priori distributions used for each parameter in the four models considered in the simulation study of Section~\ref{sec3}.

\begin{table}[H]
    \centering
    \begin{tabular}{llccccp{5cm}}
    \toprule
    \textbf{Parameter} & \textbf{Distribution} & \textbf{BSM} & \textbf{BWM} & \textbf{BCR} & \textbf{DEM} & \textbf{Description} \\ 
    \midrule
    \multicolumn{7}{l}{\textit{Fixed effects}} \\ 
    \midrule
    $\beta_0$ & $N(10,100)$ & \checkmark & \checkmark & \checkmark & \checkmark & Intercept \\
    $\beta_2$ & $N(0,100)$  & \checkmark & \checkmark & \checkmark & \checkmark & Linear slope before CP (decay rate in DEM) \\
    $\beta_{CP}$ &$U(0,20)$  & \checkmark & \checkmark & \checkmark & \checkmark & Change point \\
    $\theta_{T}$ &$U(0,5)$   &           & \checkmark  & \checkmark  & \checkmark  & Transition parameter \\ 
    \midrule
    \multicolumn{7}{l}{\textit{Random effects}} \\ 
    \midrule
    $\sigma^2_{\epsilon}$ &$HC(0,10)$  & \checkmark  & \checkmark  & \checkmark  & \checkmark  & Observation noise variance \\ 
    $\omega^2_{b_0}$       &$HC(0,1)$   & \checkmark  & \checkmark  & \checkmark  & \checkmark  & Intercept variance \\ 
    $\omega^2_{b_2}$       &$HC(0,1)$   & \checkmark  & \checkmark  & \checkmark  & \checkmark  & Slope (decay rate in DEM) variance \\ 
    $\omega^2_{b_{CP}}$    &$HC(0,1)$   & \checkmark  & \checkmark  & \checkmark  & \checkmark  & Change point variance \\ 
    \bottomrule
    \end{tabular}
    \caption{Prior distributions for the simulation study described in Section~\ref{sec3}. Note that $N$, $U$, and $HC$ denote the normal, uniform, and half-Cauchy distributions, respectively.}
    \label{priors_table}
\end{table} 
Note that $N$, $U$, and $HC$ denote the normal, uniform, and half-Cauchy distribution, respectively.

\section{MCMC diagnostics}\label{app_C}

\setcounter{figure}{0}
\renewcommand{\thefigure}{C\arabic{figure}}

\begin{figure}[H]
    \centering
    \includegraphics[width=0.8\textwidth]{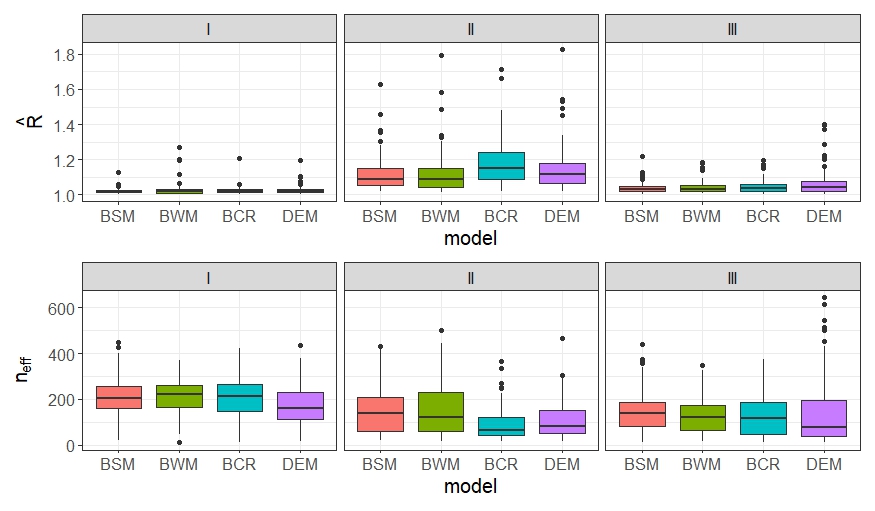}
    \caption{Convergence diagnostics for the $\hat{\beta}_{CP}$ estimate.}   \label{fig:mcmc_diag}
\end{figure}

Figure~\ref{fig:mcmc_diag} presents a summary of convergence indicators related to the $\hat{\beta}_{CP}$ estimate for the four considered models across the three considered scenarios.

\end{appendix}


\bibliographystyle{tfs}
\bibliography{bibliography.bib}

\end{document}